# Assessment of High-Frequency Performance Limits of Graphene Field-Effect Transistors

Jyotsna Chauhan and Jing Guo

Department of Electrical and Computer Engineering

University of Florida, Gainesville, FL, 32611-6130

ABSTRACT

High frequency performance limits of graphene field-effect transistors (FETs) down to a channel length of 20nm are examined by using self-consistent quantum simulations. The results indicate that although Klein band-to-band tunneling is significant for sub-100nm graphene FET, it is possible to achieve a good transconductance and ballistic on-off ratio larger than 3 even at a channel length of 20nm. At a channel length of 20nm, the intrinsic cut-off frequency remains at a couple of THz for various gate insulator thickness values, but a thin gate insulator is necessary for a good transconductance and smaller degradation of cut-off frequency in the presence of parasitic capacitance. The intrinsic cut-off frequency is close to the LC characteristic frequency set by graphene kinetic inductance and quantum capacitance, which is about $100 GHz \cdot \mu m$ divided by the gate length.

# I. Introduction

Graphene has emerged as one of the most promising materials in fundamental research and engineering applications [1-3]. Two-dimensional (2D) nature and linear bandstructure of graphene proved an ideal platform for exploring phenomenon of relativistic physics [1, 2]. Extraordinary electronic transport properties like high mobility and high saturation velocity make it attractive for radio frequency (RF) electronics applications [4-7]. Although the zero bandgap of 2D graphene leads to a low on-off ratio not desired for digital electronics applications, RF electronics applications do not require a large on-off ratio. Scaling down the channel length plays a critically important role in boosting the RF performance of a field-effect transistor (FET), and aggressive channel length scaling of graphene FET has been experimentally pursued [5-9]. Recent experiments have demonstrated graphene transistors with intrinsic cut-off frequency projected to be at the hundreds of GHz range at sub 100nm channel scale [6-7]. Fabrication of graphene transistors with the projected cut off frequency of 300 GHz for 140nm channel [8] and 100 GHz for 240nm channel [5], which significantly outperform the conventional silicon MOSFETs [10], has already been demonstrated . The issues of ultimate channel scaling and performance limits of graphene RF transistors, however, remain unclear.

In this work, we examine the RF performance limits and channel length scaling of graphene FETs with a channel length down to 20nm using self-consistent ballistic quantum transport simulations with the non-equilibrium Green's function (NEGF) formalism [11]. Quantum NEGF simulation models the Klein band-to-band (BTB) tunneling in graphene FETs, and shows that tunneling current component consists a significant fraction of the total current in graphene FETs at a short channel length below 100nm. With the channel length scaling down to 20nm, there is no significant increase in the tunneling component contribution as part of the minimal leakage current. For channel length scaling at a fixed gate insulator thickness, the decrease of the on-off ratio and transconductance for the 20nm graphene FET is attributed to electrostatic short channel effects. Thus even at a channel length of 20nm, scaling down the gate insulator thickness can improve the ballistic on-off ratio above 3. The intrinsic cut-off frequency of 20nm graphene FETs remains at a couple of THz even with a worse gating, but the extrinsic cut-off frequency degrades much less if a thin gate insulator is used. We also examine the kinetic inductance of 2D graphene and discuss possible impacts of non-quasi-static effects. The results of analyzing the high frequency behavior of graphene FETs in sub 100nm range can provide design guidelines to attain the maximum potential of graphene in the field of ultra high-speed RF devices and circuits.

## II.     Simulation Approach

Top gated graphene FETs as shown in Fig. 1 were simulated. The metal source and drain contacts are connected to the intrinsic two-dimensional channel. The nominal device has a top gate insulator thickness of $t_{ins}$=16.4nm and dielectric constant of $\kappa_{ins}$=9, which results in a gate insulator capacitance of $C_{ins}$≈486nF/cm$^2$ close to the value in a recent experiment [6]. The dielectric constant used here is close to that of GaN or Al$_2$O$_3$, which has been explored as the gate insulator for graphene FETs. The channel length is $L_{ch}$=100nm, with a zero gate underlap which could be achieved by a self align process [6]. The difference between the metal Fermi energy level and the Dirac point of graphene is $E_F$-$E_D$=-0.2eV. This contact barrier height is typical for a high work function metal contact (such as Pd) which makes a better contact for hole conduction. The device parameters mentioned here are the nominal ones, and we vary these parameters for exploring various device issues.

Graphene FETs were simulated by solving the quantum transport equation using the non-equilibrium Green's function (NEGF) formalism with the Dirac Hamilton, self-consistently with a two-dimensional Poisson equation. To assess the performance limits, ballistic transport was assumed. The Dirac Hamiltonian was discretized using the finite difference method along the carrier transport direction, defined as *y* direction,

$$H_D(k_x) = \hbar v_F \begin{bmatrix} 0 & k_x - i\hat{k}_y \\ k_x - i\hat{k}_y & 0 \end{bmatrix}, \qquad (1)$$

where $\hbar$ is the reduced Planck constant, $v_F \approx 9.3 \times 10^7 cm/s$ is the Fermi velocity, and $k_x$ is the wave vector in the transverse direction. The transverse modes are decoupled in the ballistic transport limit. For a specific transverse mode $k_x$, the Green's function was computed as,

$$G(E, k_x) = [(E + i0^+)I - H_D(k_x) - U - \Sigma_S - \Sigma_D]^{-1}, \qquad (2)$$

where *U* is the self-consistent potential, and $\Sigma_S(\Sigma_D) = -i\Delta$ and $\Delta = \pi t^2 D_0$ is the source (drain) contact self-energy of the metal contacts [12]. Here *t* is the coupling parameter between metal and graphene and $D_0$ is the metal density-of-states near its Fermi level. We use a value of $\Delta \approx 2.5 eV$ here. This phenomenological model has been extensively used before to model carrier injection from metal contacts to carbon nanotubes.

After the Green's function is calculated, the electron and hole densities are computed as,

$$n(y) = \frac{4}{W} \sum_{k_x} \int_{E_D}^{+\infty} dE[D_S(E, k_x)f_0(E - E_{FS}) + D_D(E, k_x)f_0(E - E_{FD})],$$

$$p(y) = \frac{4}{W} \sum_{k_x} \int_{-\infty}^{E_D+\infty} dE[D_S(E, k_x)(1 - f_0(E - E_{FS})) + D_D(E, k_x)(1 - f_0(E - E_{FD}))], \qquad (3)$$

where *W* is the channel width, $f_0$ is the Fermi-Dirac function, $E_{FS}$ ($E_{FD}$) is the source (drain) Fermi energy level, and $D_S$ ($D_D$) is the local-density-of-states due to the source (drain) contacts as computed by the NEGF formalism. The factor of 4 counts for a valley degeneracy of 2 and a spin degeneracy of 2.

To compute the self-consistent potential, a 2D Poisson equation is solved in the cross section as shown in Fig. 1. The potentials at source/drain and gate electrodes are fixed as the boundary conditions, and the gate flat band voltage was assumed to be zero for simplicity. (In practice, it would depend on the gate workfunction.) The iteration between the Dirac quantum transport equation and the Poisson equation continues until self-consistency is achieved, then the source-drain ballistic current is computed by

$$I = \frac{4e}{h} \int dE \cdot T(E)[f_0(E - E_{FS}) - f_0(E - E_{FD})], \tag{4}$$

where $T(E)$ is the source-drain transmission calculated by the NEGF formalism.

A quasi-static treatment was used to assess high-frequency performance of graphene FETs [13, 14]. The intrinsic gate capacitance, $C_g$ and the transconductance, $g_m$, are computed by running the above self-consistent DC simulations at two slightly different gate voltages and computing derivatives of the charge in the channel and the drain current numerically,

$$C_g = \left.\frac{\partial Q_{ch}}{\partial V_G}\right|_{V_D}, \quad g_m = \left.\frac{\partial I_D}{\partial V_G}\right|_{V_D}, \tag{5}$$

The intrinsic cut-off frequency is computed as,

$$f_T = \frac{1}{2\pi} \frac{g_m}{C_g}. \tag{6}$$

The parasitic capacitance exists between the gate and source (drain) electrode as $C_{ps}(C_{pd})$. The parasitic capacitance is gate-voltage independent and the total gate parasitic capacitance is $C_p = C_{ps} + C_{pd}$. The extrinsic cut-off frequency is computed as,

$$f_T = \frac{1}{2\pi} \frac{g_m}{C_g + C_p}. \tag{7}$$

### III. Results and Discussions

It is important to scale down the channel length for boosting the RF performance of graphene FETs. Recent experiments on sub-100nm graphene FETs mostly focus on graphene FETs with a channel length between 50nm and 100nm. Because graphene is a zero band gap material where Klein band-to-band (BTB) tunneling plays an important role [15], it might be expected that the off current can significantly increase resulting in a lack of gate modulation as the channel length further scales down. Previous modeling work of graphene FETs showed a large leakage current due to tunneling [16]. Except examining channel length scaling down to 20nm and BTB tunneling, the reminder of the result section also addresses the issues of electrostatic design of 20nm graphene FETs for RF applications, and importance of non-quasi-static effect and kinetic inductance of graphene as the device intrinsic cut-off frequency approaches THz regime.

We start by simulating the ballistic I-V characteristics of a graphene FET with a channel length of 100nm, as shown in Fig. 2. Figure 2a shows an asymmetric $I_D$-$V_G$ characteristic for electron and hole conductions, because the metal contacts make better contacts for holes compared to electrons. The maximum transconductance simulated at the ballistic limit is $g_m \approx 1310 \mu A/\mu m$ at a drain voltage of $V_D$=-0.5V. A maximum transconductance of $g_{m,exp} \approx 1100 \mu A/\mu m$ was reported in a recent experiment for a self-aligned graphene FET with the same gate capacitance and channel length at $V_D$=-0.5V [6]. The closeness of the experimental value of the transconductance to the simulated ballistic value indicates the high quality of graphene channel and low parasitic resistance for the experimental device. Figure 2b shows the simulated $I_D$ vs. $V_D$ characteristics. For the simulated bias regimes, the source-drain ballistic current increases approximately linearly with the applied drain bias voltage, which qualitatively agrees with the measured data on the 100nm graphene FET reported by Liao et al.[6]. To examine the issue of ultimate channel length scaling for boosting the performance of graphene FETs, we simulated the I-V characteristics of graphene FETs of different channel lengths down to 20nm as shown in Fig. 3. The nominal values of the gate insulator thickness and dielectric constant are used. As shown in Fig. 3a, the minimal leakage current increases as the channel length scales down below 100nm. The increase is especially considerable as the channel length decreases from 30nm to 20nm. On the other hand, the on-current (which is defined as the current at $V_G$=-2.25V) remains almost constant as the channel length scales down to 40nm, and it decreases as the channel length further scales to 20nm. The left axis of Fig. 3b plots the on-off ratio as a function of the channel length. The value is slightly larger than 2 for a channel length of 100nm, and decreases to a value of 1.25 for a channel length of 20nm. Since it is somewhat arbitrary to choose the gate voltage at which the on-current is defined, it is useful to compare the transconductance. The right axis of Fig. 3b shows the transconductance as a function of the channel length. As the channel length scales from 100nm to 40nm, the ballistic transconductance only decreases very slightly from the value of about 1300µS/µm. As the channel length scales down to 20nm, the transconductance, however, drops significantly to about 346 µS/µm, which is only about 26% of the value at the channel length of 100nm. Since the cut-off frequency is proportional to transconductance, significant lowering of transconductance is not preferred.

The dependence of the cut-off frequency on the channel length is examined next. As shown in Fig.4, the intrinsic cut-off frequency keeps increasing as the channel length scales from 100nm down to 20nm, although the transconductance decreases considerably as the channel length scales down from 40nm to 20nm as described before. The simulated intrinsic $f_T$ is about 640GHz at $L_{ch}$=100nm and about 1.37THz at $L_{ch}$=50nm. To understand this result, we plotted the intrinsic gate capacitance (per unit channel width) as a function of the channel length as shown in Fig. 4b. The gate capacitance decreases approximately linearly as the channel length decreases due to two reasons. First, a smaller channel length results in a smaller gated channel area per unit channel width. Second, as the channel length decreases, electrostatic short channel effects become important, especially when the channel length becomes comparable to the gate insulator thickness. The gate modulation of the channel potential and charge becomes less effective. The gate capacitance, therefore, decreases. The decrease of the gate capacitance outpaces the decrease of the transconductance. The intrinsic cut-off frequency monotonically increases as the channel length decreases from 100nm to 20nm. We also noticed that intrinsic cut off frequency decreases at large gate drive voltages due to considerable population of –$k$ states resulting in decrease of the average carrier velocity. This phenomenon has already been reported for carbon nanotube FETs [14].

If a parasitic capacitance is considered, the extrinsic cut-off frequency decreases below $L_{ch}$~40nm due to significant decrease of $g_m$. Since the parasitic capacitance plays an increasingly important role as the channel length decreases due to a larger parasitic to intrinsic capacitance ratio, it is important to maintain a large enough transconductance to ensure good high frequency performance at a short channel length.

The degradation of the transconductance at short channel lengths as shown in Fig. 3b could be due to either carrier transport effects or transistor electrostatic design. For carrier transport, it might be concerned that a decrease of the channel length could result in significant Klein BTB tunneling and thereby less effective gate modulation. To examine this effect, we plot the potential profile and the energy-resolved current spectrum for the modeled 50nm graphene FET as shown in Fig. 5a. The non-tunneling and the Klein BTB tunneling current can be identified as follows. The current delivered in the energy range below the minimum value of the Dirac point in the channel, as shown by the dashed line in Fig. 5a, is identified as non-tunneling current because a carrier always remains in the valence band as it travels from the source to drain. In contrast, in the energy range between the minimum value and the maximum value of the Dirac point in the channel, the current is identified as the tunneling component because a carrier goes between the conduction band and the valence band by Klein BTB tunneling as it travels from the source to drain. Figure 5b plots the percentage of the BTB tunneling current in the total drain current as a function of the channel length. Although the BTB tunneling component is significant and can account for over one half of the total drain current, the percentage of the BTB tunneling current in the total drain current remains almost constant as the channel length decreases from 50nm to 20nm. It indicates that the increase of the tunneling component in the total current is so slight that it cannot be responsible for the significant decrease of the transconductance as the channel length decreases from 50nm to 20nm.

To examine transistor electrostatic effect and optimize electrostatic design, we simulated $I_D$ vs $V_G$ characteristics by decreasing the gate insulator thickness, while the channel length is fixed at $L_{ch}$=20nm, as shown in Fig. 6. Significantly improved gate modulation and a larger transconductance are observed in Fig.6a, especially when the gate insulator thickness decreases below 5nm. It indicates that the thick gate insulator compared to the short channel length of 20nm, is mostly responsible for the degradation of the transconductance at a channel length of 20nm. Figure 6b shows the intrinsic and extrinsic cut-off frequencies as a function of the gate insulator thickness for the 20nm graphene FET. The intrinsic cut-off frequency reaches a peak value of about 3.7THz at a gate insulator thickness of 8nm. The decrease of the intrinsic cut-off frequency as $t_{ins}$ decreases below 8nm is due to the increase of the gate capacitance, which is the serial combination of the insulator capacitance and the graphene quantum capacitance. As the gate insulator becomes thin, the gate modulation is more effective and carrier populates energy ranges with higher density-of-states which results in an increase of the quantum capacitance. If a parasitic capacitance with a value close to the intrinsic gate capacitance at $t_{ins}$=2nm is considered as shown by the dashed line in Fig. 6(b), the extrinsic cut-off frequency at $t_{ins}$=2nm significantly outperforms that at $t_{ins} \approx$16nm.

Next we compute the kinetic inductance of graphene by extending the derivation of kinetic inductance of carbon nanotubes [17]. If the +k states are filled by $E_{FS} = eV/2$ and -k states are filled by $E_{FD} = -eV/2$, the current is $I = We^3V^2/(8\pi^2\hbar^2 v_F)$ per valley per spin. The net increase of the energy of the system is computed as the excess energy of moving carriers from the valence band to the conduction band $E_k = WL_{ch}e^3V^3/(48\pi\hbar^2 v_F^2)$. Since $dE_k = d(\frac{1}{2}L_k I^2)$, the kinetic inductance is

$$L_k = \frac{1}{I}\frac{dE_k}{dI} = \frac{2\pi^3\hbar^2}{e^3 V}\frac{L_{ch}}{W} = \frac{\pi^3\hbar^2}{e^2 E_F}\frac{L_{ch}}{W} \approx 84pH \times \left(\frac{1eV}{E_F}\right)\left(\frac{L_{ch}}{W}\right). \tag{8}$$

The kinetic inductance is plotted as a function of the Fermi level in Fig. 7a. It is inversely proportional to the Fermi energy because the number of transverse modes linearly increase as a function of the Fermi energy. For a Fermi level $E_F = eV/2$, the total equilibrium charge at zero temperature is $Q = \frac{WL_{ch}e(eV/2)^2}{4\pi\hbar^2 v_F^2}$ per spin per valley. The quantum capacitance is expressed as,

$$C_Q = \frac{dQ}{dV} = \frac{e^3 V}{8\pi\hbar^2 v_F^2} WL_{ch} \qquad (9)$$

The LC characteristic frequency is

$$f_{LC} = \frac{1}{2\pi}\frac{1}{\sqrt{L_k C_Q}} = \frac{1}{2\pi}\frac{(2/\pi)v_F}{L_{ch}} \approx \frac{100 GHz \cdot \mu m}{L_{ch}}, \qquad (10)$$

where the gate length $L_g = L_{ch}$ for the simulated device. Equation 10 indicates that the LC characteristic frequency is proportional to an average velocity of $(2/\pi)v_F \approx 6 \times 10^7$ cm/s, which can be interpreted as the average velocity along the transport direction for a 2D graphene with $+k$ states populated. Figure 7b plots the LC characteristic frequency and compare it to the simulated cut-off frequency as a function of the channel length for sub 100nm-graphene FET. A rigorous treatment beyond quasi-static approximation requires inclusion of capacitive, resistive, and inductive elements for calculation [17, 18]. The quasi-static approximation includes the equivalent capacitive and resistive elements, but omits the equivalent inductive elements. In order to assess how good the quasi-static approximation is, one can compare the operation frequency to the LC characteristic frequency, which is about 1THz for a 100nm graphene FET. Figure 7b shows the intrinsic cut-off frequency is close to this value. The non-quasi-static effect, therefore, could be important if the graphene FET operates at its intrinsic cut-off frequency. On the other hand, the extrinsic cut-off frequency of a short channel graphene FET could be much lower than its intrinsic value if the parasitic gate capacitance is not reduced to a value comparable to the intrinsic gate capacitance. In this case, non-quasi-static effect is not important. Furthermore, ballistic transport is assumed in this study to assess the RF performance limits of short channel graphene FETs. The cut-off frequency can be lowered by scattering [19], which is beyond the scope of this work.

The intrinsic cut off frequency of ballistic carbon nanotube FETs has been reported to be in the range of 80-110GHz/$L_{ch}$(µm) [19,20]. Figure 7b shows the intrinsic cut off frequency of ballistic carbon nanotube FETs as function of channel length calculated using the relation $f_T = \frac{110 GHz}{L_{ch}(\mu m)}$.

It is observed that the cut off frequency of ballistic carbon nanotube FETs is slightly better than ballistic graphene FETs. This difference can be attributed to the following two reasons. First, since a carbon nanotube is a one dimensional material as compared to graphene which is two dimensional, averaging carrier velocity along the transport direction requires projection of the velocity along the transport direction in a 2D channel. Second, graphene has a linear E-k with a zero bandgap, as compared to carbon nanotube with a parabolic E-k with a finite bandgap. The difference in bandstructure results in different population of $-k$ states and different bandstructure-limited velocities.

**IV.    Conclusions**

In Summary, we study channel length scaling and assess RF performance limits of graphene FETs in the sub-100nm channel length regime by using self-consistent ballistic quantum transport simulations. The simulated intrinsic cut-off frequency is about 640GHz at a channel length of 100nm and increases to about 3.7THz at $L_{ch}$=20nm. For a gate insulator thickness of 16nm, scaling down the channel length to 20nm does result in significant decrease of on-off current ratio. Because of the low transconductance, the high cut-off frequency is highly susceptible to parasitic gate capacitance. As the gate insulator scales down to about 1/10 of the 20nm channel length, the on-off current ratio can increase to about 3, with a cut-off frequency much less susceptible to parasitic capacitance. To discuss the non-quasi-static effect, the kinetic inductance of the graphene is computed and the LC characteristic frequency is about $100 GHz \cdot \mu m/L_g$. As the intrinsic cut-off frequency is close to this LC characteristic frequency, we expect the non-quasi-static effects can start to play a role as the transistor is optimized to perform close to its intrinsic cut-off frequency.

**Acknowledgement**

We would like to thank Dr. Eric Snow of Naval Research Lab (NRL) for bringing our attention to this important problem, and Prof. K. Shepard and I. Meric of Columbia university for helpful discussions on Klein tunneling. This work was supported by ONR, NSF, and ARL.

# References

[1] Novoselov K. S. *et al.*, "Electric Field Effect in Atomically Thin Carbon Films," *Science* **2004**, 306, 666–669.

[2] Zhang Y. *et al.*, "Experimental observation of the quantum Hall effect and Berry's phase in graphene," *Nature* **2005**, 438, 201–204.

[3] Li X. *et al.*, "Chemically Derived, Ultrasmooth Graphene Nanoribbon Semiconductors," *Science* **2008**, 319, 1229–1232.

[4] Meric I., Baklitskaya N., Kim P., and Shepard K., "RF performance of top-gated, zero-bandgap graphene field-effect transistors," *Tech. Dig. of Int. Electron Device Meeting* (IEDM) **2008**, 4796738, 1–4.

[5] Lin Y., Dimitrakopoulos C., Jenkins K., Farmer D., Chiu H., Grill A., and Avouris Ph., "100-GHz Transistors from Wafer-Scale Epitaxial Graphene," *Science* **2010**, 327, 662.

[6] Liao L., Bai J., Cheng R., Lin Y., Jiang S., Qu Y., Huang Y., and Duan X., "Sub-100nm Channel Length Graphene Transistors," *Nano Lett*. **2010**, 10, 3952-3956.

[7] Wu Y., Lin Y., Jenkins K. et al, "RF Performance of Short Channel Graphene Field-Effect Transistor," *Tech. Dig. of Int. Electron Device Meeting* (IEDM) **2010**, 226–228.

[8] Liao L. *et al*., "High-speed graphene transistors with a self-aligned nanowire gate," *Nature* **2010**, 467, 305-308.

[9] Lin Y. *et al*., "Operation of Graphene Transistors at Gigahertz Frequencies," *Nano Lett.* **2009**, 9 (1), 422–426.

[10] Schwierz F., "Graphene Transistors," *Nature Nanotechnology* **2010**, 5, 487-496.

[11] Datta S., *Quantum transport: Atom to transistor*. Cambridge, UK: Cambridge University Press, **2005**.

[12] Svizhenko A. and Anantram M. P., "Effect of scattering and contacts on current and electrostatics in carbon nanotubes," *Phys. Rev. B* **2005**, 72, 085430.

[13] Rutherglen C., Jain D., and Burke P., "Nanotube electronics for radiofrequency applications," *Nature Nanotechnology* **2009**, 4, 811-819.

[14] Guo J., Hasan S., Javey A., Bosman G., and Lundstrom M., "Assessment of High-Frequency Performance Potential of Carbon Nanotube Transistors," *IEEE Trans. on Nanotechnology* **2005**, 4, 715-721.

[15] Katsnelson M., Novoselov K., and Geim A., "Chiral tunnelling and the Klein paradox in graphene," *Nature Physics* **2006**, 2, 620-625.

[16] Ryzhii V., Ryzhii M., and Otsuji T., "Thermionic and tunneling transport mechanisms in graphene field-effect transistors," *Phys. Stat. Sol (a)* **2008,** 205, 1527-1533.

[17] Burke P., "Luttinger Liquid Theory as a Model of the Gigahertz Electrical Properties of Carbon Nanotubes," *IEEE Transactions on Nanotechnology* **2002**, 1, 129-144.

[18] Chen Y., Ouyang Y., Guo J., and Wu T. "Time-dependent quantum transport and non-quasistatic effects in carbon nanotube transistors," *Applied Physics Letters* **2006**, 89, 203122.

[19] Yoon Y., Ouyang Y., and Guo J., "Effect of Phonon Scattering on Intrinsic Delay and Cut-Off Frequency of Carbon Nano Tube FETs," *IEEE Trans. on Electron Devices* **2006**, 53, 2467-2470.

[20] Burke P., "AC performance of nanoelectronics: towards a ballistic THz nanotube transistor," *Solid-State Electronics* **2004**, 48, 1981-1986.

# Figures

Figure 1. Modeled graphene field-effect transistor. The two-dimensional graphene channel is contacted to the metal source and drain contacts, and is modulated by the top gate.

Figure 2. Simulated I-V characteristics for the nominal device described in text. (a) The $I_D$ vs. $V_G$ characteristic at $V_D$=-0.5V. (b) The $I_D$ vs. $V_D$ characteristics at $V_G$=-0.25V to -2.25V at -0.5V per step (from the bottom to top curve). The graphene channel length is $L_{ch}$=100nm. The flat band voltage is zero. The source/drain contact barrier height for holes is $E_F - E_D = -0.2eV$, which is the difference between the metal Fermi level and the Dirac point of graphene.

Figure 3. Channel length scaling (a) $I_D$ vs. $V_G$ characteristics for the graphene FET as shown in Fig. 1 with different channel lengths, $L_{ch}$=100, 50, 40, 30 and 20nm. (b) The on-off current ratio (left axis) and transconductance (right axis) as a function of the channel length. The on-current is computed at $V_G$=-2.25V and the off-current is at $V_G$=0V. The transconductance is obtained at $V_G$=-1.5V. The top gate insulator thickness is $t_{ins}$=16nm and dielectric constant is $\kappa_{ins}$=9. The applied drain voltage is $V_D$=-0.5V.

Figure 4. (a) Intrinsic (solid) and extrinsic (dashed) cut-off frequency as a function of the channel length for the graphene FET as shown in Fig. 1. (b) The intrinsic gate capacitance as a function of the channel length. The top gate insulator thickness is $t_{ins}$=16nm and dielectric constant is $\kappa_{ins}$=9. The applied gate voltage is $V_G$=-1.5V and drain voltage is $V_D$=-0.5V. A constant parasitic capacitance of $C_p = 500 aF/\mu m$ is assumed for computing the extrinsic cut-off frequency.

Figure 5. Band-to-band tunneling (a) The Dirac point of graphene as a function of the channel position (bottom axis) and the current spectrum (top axis) for the graphene FET as shown in Fig. 1 with a channel length of $L_{ch}$=50nm at $V_G$ =-1.5V and $V_D$ =-0.5V. The current component below the dashed is identified as non-tunneling component because carriers always remain in the valence band as they travel from source to drain. The current component above the dashed line is identified as Klein band-to-band tunneling component because carriers in the conduction band near the source go to the valence band as they travel from source to drain. (b) Ratio between the BTB tunneling current and the drain current as a function of the channel length at $V_G$ =-0.25V (solid), -1.5V (dashed) and $V_D$ =-0.5V.

Figure 6. (a) $I_D$ vs. $V_G$ characteristics for a graphene FET as shown in Fig. 1 with a channel length of $L_{ch}$=20nm and different gate insulator thicknesses, $t_{ins}$=16nm (cyan solid),12nm(pink with asterisks), 8nm(black with triangles), 5nm(red with squares) and 2nm(blue with circles). The applied drain voltage is $V_D$=-0.5V. (b) The intrinsic (solid) and extrinsic (dashed) cut-off frequency as a function of the top gate insulator thickness. The cut-off frequencies are computed at $V_G$ =-1.25V and $V_D$ =-0.5V. A constant parasitic capacitance of $C_p = 100aF/\mu m$ (dashed pink with diamonds), $C_p = 500aF/\mu m$ (dashed red with squares) and $C_p = 1000aF/\mu m$ (dashed black with triangles) is assumed for computing the extrinsic cut-off frequency.

Figure 7. (a) Kinetic inductance as a function of the Fermi energy level for 2D graphene. (b) The LC characteristic frequency (blue solid line) ,the intrinsic cut-off frequency $f_T$ (red dashed line) of graphene FETs and the intrinsic cut-off frequency $f_T$ (black dashed-dotted line) of ballistic carbon nanotube FETs [19] as a function of the channel length.

The intrinsic cut-off frequency is computed at $V_G$=-1.5V and $V_D$=-0.5V. The top gate insulator thickness is $t_{ins}$=16nm and dielectric constant is $\kappa_{ins}$=9.

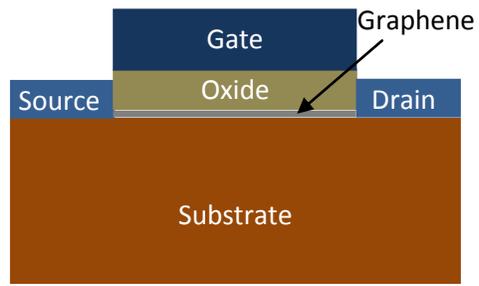

Figure 1

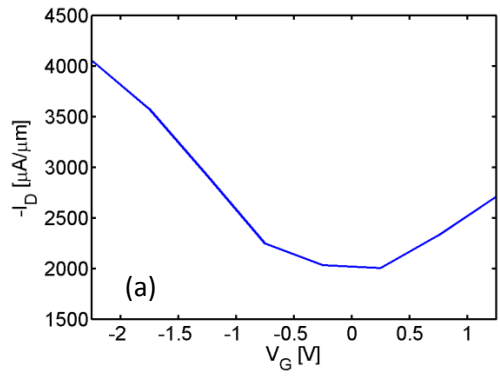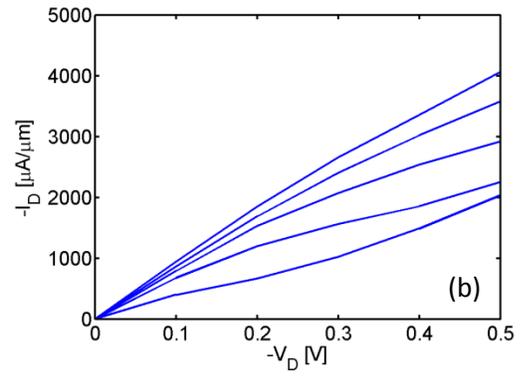

Figure 2

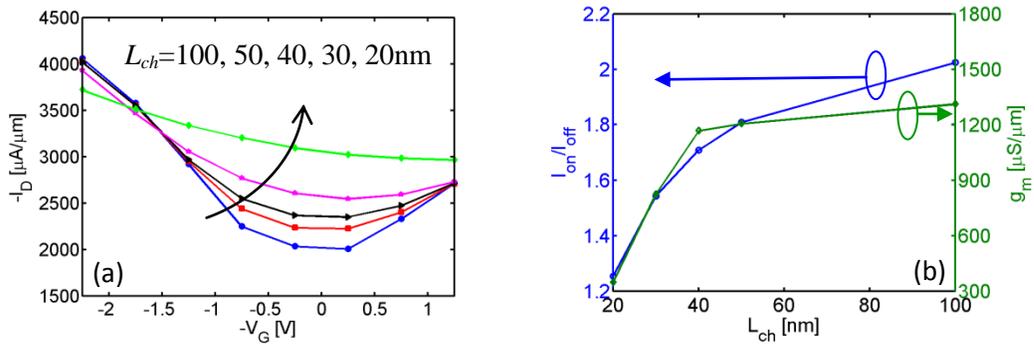

Figure 3

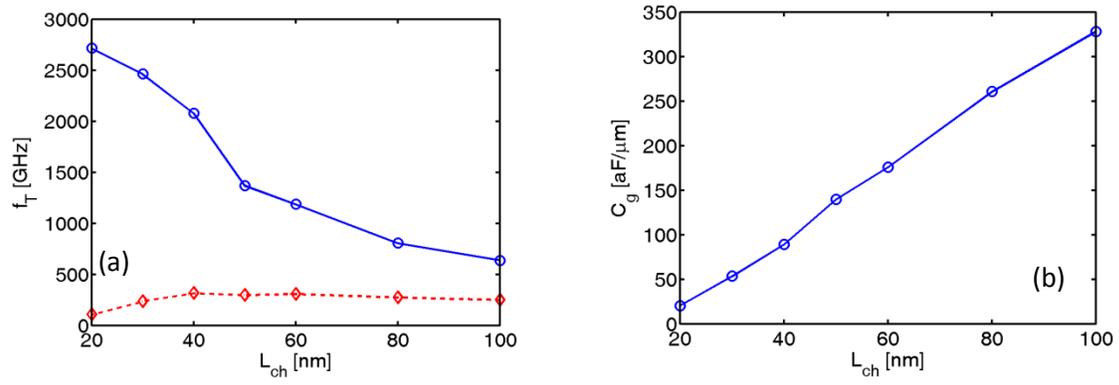

Figure 4

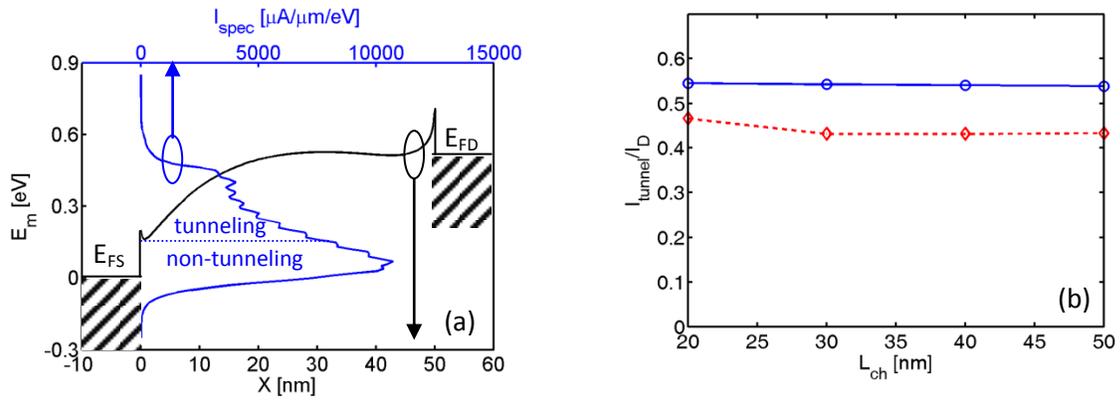

Figure 5

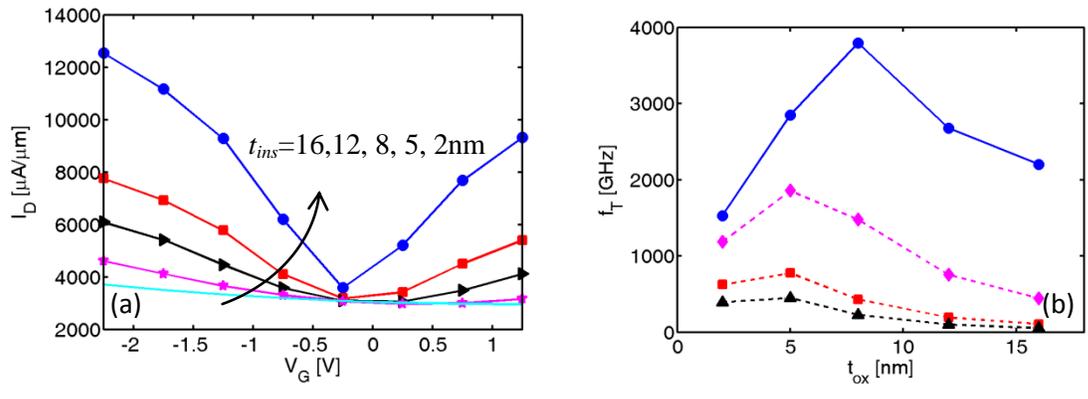

Figure 6

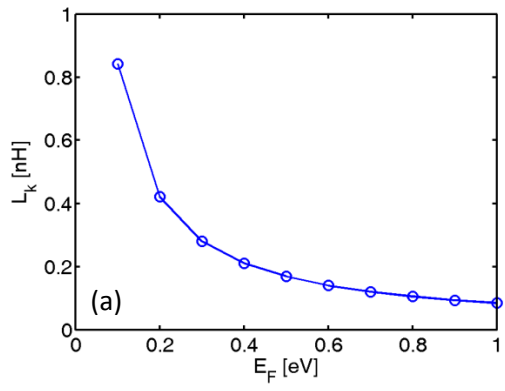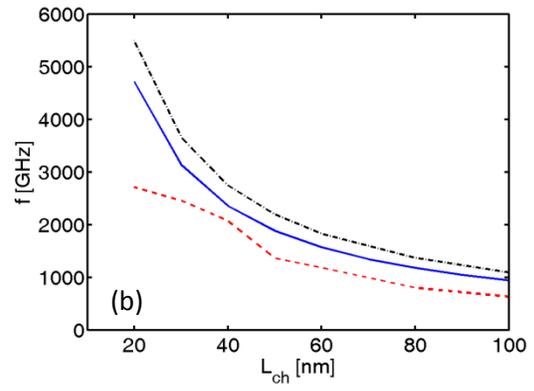

Figure 7